\documentclass[apj]{emulateapj}

\usepackage{url,amsfonts,epsfig}
\usepackage{graphicx}
\usepackage{amsmath}
\usepackage{amssymb}
\usepackage[section]{placeins}
\usepackage{chngpage}
\usepackage{calc}
\usepackage{subfigure}
\usepackage{mathtools}

\usepackage{natbib}
\makeatletter 
\renewcommand\@biblabel[1]{}

\shorttitle{Nuclear Star Clusters around Massive Black  Holes}
\shortauthors{Antonini}
\begin{document}
\def\gap{\;\rlap{\lower 2.5pt
\hbox{$\sim$}}\raise 1.5pt\hbox{$>$}\;}
\def\lap{\;\rlap{\lower 2.5pt
 \hbox{$\sim$}}\raise 1.5pt\hbox{$<$}\;}

\newcommand\sbh{MBH}
\newcommand\NSC{NSC}
\title{Origin and growth of nuclear star clusters around massive black  holes}
%\title{Dissipationless formation and Evolution of the Milky Way Nuclear Star Cluster. II.~the Galactic center distribution of stellar remnants. }

\author{Fabio Antonini}
\email{antonini@cita.utoronto.ca}
\affil{Canadian Institute for Theoretical Astrophysics, University of Toronto,
60 St. George Street, Toronto, Ontario M5S 3H8, Canada}

\begin{abstract}
The centers of stellar spheroids less luminous than $\sim 10^{10}~L_{\odot}$ are often 
marked  by the presence of nucleated central regions, 
called  ``nuclear star clusters''~(\NSC s). The origin  of \NSC s is still unclear.
Here we investigate  the possibility  that  \NSC s originate from the 
migration  and merger of stellar clusters  
 at the center of galaxies where a  massive black hole~(\sbh) may sit.
We show  that the observed scaling relation between \NSC\ masses and the 
velocity dispersion of their host spheroids cannot be reconciled with a 
purely ``in-situ'' dissipative formation scenario. On the other hand,
the observed relation appears to be in agreement with the predictions of 
the cluster merger  model. A dissipationless  formation model
also reproduces the observed relation between 
the size of \NSC s and their total luminosity, 
$R\propto \sqrt{\mathcal{L}_{\rm NSC}}$. When a  \sbh\ is included at the center of the galaxy,
such   dependence becomes substantially 
 weaker than the observed correlation, since the size of the \NSC\ is 
mainly determined by the fixed tidal field of the \sbh . 
We evolve  through dynamical friction a population of stellar clusters in a model of a galactic bulge
taking into account dynamical dissolution due to two body relaxation,
starting   from  a power-law cluster initial mass function~(CIMF) 
and adopting an initial total mass in stellar clusters
consistent with the present-day cluster formation efficiency
of the Milky Way~(MW).
The most massive clusters reach the center of the galaxy and merge  to
form a compact  nucleus; after $10^{10}$ years,
the resulting \NSC\ has properties 
that are consistent with the observed distribution  of stars  in the MW \NSC .
When a \sbh\ is included at the center of a galaxy, globular 
clusters are tidally disrupted during inspiral, resulting in  \NSC s with  
lower  densities  than those of \NSC s forming in galaxies with no \sbh s. 
We suggest  this as a possible explanation for  
the lack of  \NSC s  in galaxies containing \sbh s more massive 
than $\sim 10^8~M_{\odot}$.
Finally, we investigate the orbital evolution of globular clusters in giant  elliptical  
galaxies which are  believed  to always host a \sbh\ at their center rather than a \NSC . 
In these systems an additional mechanism can prevent a \NSC\ from forming:
the time for globular clusters to reach the center of the galaxy is  much longer than the
Hubble time. 
  \end{abstract}
 

\section{introduction}
Many galaxies, over the whole Hubble sequence, show nucleated central 
regions often referred to as  ``nuclear star clusters''~(\NSC s). These systems  are  among the densest star 
clusters observed, with effective radii of a few parsecs and central luminosities 
up to $\sim 10^7~ L_{\odot}$~\citep{MG97,carollo2,carollo,BSM:02,Balcells+07,GG03,BSM:04,Cote,turner+12}. 
The total frequency of nucleation  is as large as  $80\%$ for galaxies fainter than $M_B=-19.5$, while 
\NSC s tend to disappear in galaxies brighter than this magnitude.

There is no consensus on how \NSC s form. A \emph{dissipative}  origin bases on the 
hypothesis of radial gas  inflow into the galactic center and requires 
efficient dissipation mechanisms to work~\citep[e.g.,][]{LT82}.
In this model, a \NSC\ consists mostly of stars that formed locally \citep{S06,Sch08,Milos04,EV:08,SB:89,B07}. 

Alternatively,  \NSC s could have a  \emph{dissipationless} origin in which  massive stellar~(globular-like) 
clusters migrate to the center due to dynamical friction and  merge to form a dense nucleus 
growing up to the size of observed \NSC s~\citep{TOS}. 
Observations 
of NSCs in dE galaxies suggest that the majority of these systems could be the
result of accumulating  mass in the form of orbitally decayed globular clusters. 
Numerical studies have also shown that such a formation model  is consistent with the measured sizes and luminosities
of nuclei~\citep{CD:93,B04,CDM08,Hartmann}.  In addition, stellar population synthesis studies of NSC spectra suggest  that most of the mass usually resides
in stars as old as typical  globular clusters~\citep[$\sim 10~$Gyr;][]{Figer2004,BKR10}. On the other hand, the observed correlation between colors and luminosities 
together with the complex star formation histories that often characterize the central region of galaxies,
may be difficult to explain on the basis of a dissipationless origin, unless there is some  contribution from continuous or recurrent star formation in addition to  the ancient globular cluster stars~\citep{AM12-2}. 
This suggests  that dissipation and dissipationless processes
 are not exclusive, and \NSC s  might indeed originate from a combination of the two processes.

Due to their small sizes and their crowded stellar fields,  galactic nuclei beyond
the Local Group  are typically unresolved, and the only quantities 
that can be determined are integrated properties such as half-mass radius and total luminosity. 
A radial density profile and velocity structure can be reliably determined only for the Milky Way~(MW)
\NSC~\citep{Genzel03,ShEc,GS:09,Oh}, which, due to its proximity~($\sim 8~$kpc), 
 can be resolved into individual stars~\citep{ShEc,S09}. 
 In addition to the MW, NGC~205 also has a spatially-resolved \NSC~
\citep[Figure~2 of][]{M09}.
The MW  NSC has an estimated mass of $\sim 10^{7}M_{\odot}$~\citep{LZM,S08}
and hosts a massive black hole~(\sbh )  whose mass, $\sim 4.3\times 10^6 M_\odot$,
 is uniquely well determined~\citep{Genzel03,Ghez08,Gil}. 
A number of other galaxies also contain both a \NSC  \ and a \sbh \ \citep{seth,GD:07,GS:09}, 
that have  comparable masses.
In models of NSCs,  the dynamical  influence of a \sbh \ 
should therefore be considered, at least in bulges brighter than about $10^{9}L_{\odot}$
which are believed to always contain  a \sbh\ \citep{FF:05}. 

More recently, we have presented large-scale $N$-body simulations of the inspiral 
and merger of massive clusters in the inner regions of the Galaxy~\citep{AM12-2}. We showed that
current observational constraints are consistent with the hypothesis that a large fraction of the MW NSC 
mass is in old stars brought in by infalling globular clusters.
In this paper, we expand on this previous work 
and use  simple analytical considerations to investigate
 the possibility that globular clusters can migrate through dynamical friction in the 
center of galaxies and form compact stellar nuclei. In particular, we focus on how  the presence of \sbh s
at the center of galaxies can impact the merger hypothesis for the formation of their \NSC .
In order to highlight the role of \sbh s in \NSC\ formation, 
we will systematically present our results for both cases of galaxy with and without central \sbh .

%We stress here that, due to the difficulties of resolving the  sphere of influence of \sbh s
%less massive than $\sim 10^6~M_{\odot}$, the existence of  \sbh s in most intermediate and low luminosity galaxies 
% remains speculative~\citep{Merritt-book}. 

This paper is organized as follows.
We start in \S2 by comparing  some of the observed scaling relations for \NSC s with the same  relations  predicted 
in  both  the merger and  the gas model.
In \S3 and \S4 we develop a simple analytical model  for studying the 
orbital decay of globular clusters in galaxies like the MW, and explore the effect  that a 
central \sbh\ in the galaxy   has on the properties of the resulting \NSC. In \S5 we 
use an approach similar to that of \S3 to follow
 the orbital evolution of globular clusters near massive \sbh s in the central regions of bright elliptical galaxies.
In \S6 we discuss a variety of astrophysical implications of a 
dissipationless origin of \NSC s
and conclude in \S7.

  \section{Scaling relations}
 The study of \NSC s  has revealed  a number of correlations 
 between their masses   and several global 
properties of their host galaxies, such as  velocity dispersion and bulge mass.
The existence of such correlations  might indicate a direct link 
among large galactic spacial  scales and the much smaller scale of the nuclear environment. 
While it was for long believed that \NSC s and \sbh s followed similar scaling relations with 
their host galaxies~\citep{F:06,WH:06}, 
it is now well established  that NCs do not follow any of the scaling relations defined 
by \sbh s~\citep{Balcells+07,GS:09,G12,SG12,Nathan}.
Observations also reveal that, unlike globular clusters, \NSC\ half-light
radii are luminosity dependent,  increasing  with increasing total mass~\citep{Cote,FORBES}.
This relation might contain important information on the  processes that shaped the
central regions of galaxies and their  \NSC s. 

We start here by comparing such observed correlations 
with predictions from both  the dissipative and  dissipationless  formation models.
  
 \subsection{$M_{\rm NSC} - \sigma$ relation}
Of the global-to-nucleus relations, the most frequently referred  to is the tight correlation between 
\NSC\ mass, $M_{\rm NSC}$, and  the host galaxy's velocity dispersion, 
$\sigma$. 
\citet{F:06}  argued that
such  a correlation has a slope which is consistent with that of  the well-known $M_{\bullet}-\sigma$ relation
obeyed by the  \sbh\ mass, $M_{\bullet}$.
More recently, a series of papers reached a different conclusion, suggesting   that the
\NSC\ scaling relations are instead substantially 
shallower than the corresponding \sbh \ scaling relations~\citep{GS:09,G12,SG12,Nathan}.
The  version of the  $M_{\rm NSC}-\sigma$ relation given in \citet{G12} is
\begin{eqnarray}
\log\left(\frac{M_{\rm NSC}}{M_{\odot}}\right) &=&(6.83\pm0.07) \\
&+&(1.57\pm0.24)\log(\sigma/70 {\rm km~s^{-1}})~.~~~ \nonumber
\end{eqnarray}

We mention  here  that the $M_{\rm NSC}-\sigma$ relation  
might be a non-primary correlation, instead  resulting  from  a projection of the fundamental plane,
given the observed  correlation between  $M_{\rm NSC}$ and
the total host galaxy mass~($M_{\rm gx}$): $M_{\rm NSC}\propto M_{\rm gx}^{1.18\pm 0.16}$~\citep[e.g., see][]{Nathan}.

 \subsubsection{Dissipationless model}
 A predicted $M_{\rm NSC}-\sigma$ relation can be easily derived in the globular cluster merger model
 if we  assume that  the globular cluster distribution initially follows the stellar light
  and by using an  isothermal sphere density model: $\rho(r)=\sigma^2/2\pi G r^2$,
 where $\sigma$ is the 1D velocity dispersion and $G$ the gravitational constant. The cumulative mass within $r$ is 
  $M(r)=rv_c^2/G=2r\sigma^2/G$, with $v_c$ the velocity of a circular orbit. 
The dynamical friction coefficient for a stellar cluster of mass $m_{\rm cl}$ moving
in an isothermal sphere model is~\citep{CH:43}:
 \begin{equation}\label{eq1}
\boldsymbol{f}_{\rm df} = -4\pi G^2 m_{\rm cl} \rho(r)  {{\boldsymbol v} \over v^3 }
\ln \Lambda  \left[{\rm erf}(X)-\frac{2X}{\sqrt\pi} e^{-X^2}\right]~,
 \end{equation}
 with $\ln \Lambda$ the Coulomb logarithm and $X=v/\sqrt{2}\sigma$.
 Noting that
 \begin{equation}\label{eqam}
\frac{1}{r}{dr \over dt}=-|\boldsymbol{f}_{\rm df}| \left({{dL}\over{dr}}\right)^{-1},
 \end{equation}
where $L(=\sqrt{GM(r)r})$ is the orbital angular momentum, for a cluster moving on a circular orbit we obtain
 \begin{equation}\label{eq3}
\frac{dr}{dt}=-Gm_{\rm cl} {\ln \Lambda  \left[{\rm erf}(X)-\frac{2X}{\sqrt\pi} e^{-X^2}\right] }/\sqrt{2}r\sigma~.
\end{equation}
Integrating equation~(\ref{eq3}) yields
\begin{equation}\label{eq4}
r_{in}=\left(\sqrt{2}G m_{\rm cl} \ln \Lambda   \left[{\rm erf}(X)-\frac{2X}{\sqrt\pi} e^{-X^2}\right] t/\sigma \right)^{1/2}~.
\end{equation}
Clusters initially within $r_{in}$ reach the center  at a time $\leq t$.

Assuming that the total mass accumulated in the center is equal to the total 
mass in globular clusters that are initially within $r_{in}$, then we obtain~\citep[e.g.,][]{TOS}
\begin{eqnarray}\label{eq5} 
M_{\rm NSC}&=&2^{5/4}  \left(\ln \Lambda  \left[{\rm erf}(X)-\frac{2X}{\sqrt\pi} e^{-X^2}\right] \langle m_{\rm cl} \rangle t/G \right)^{1/2} \nonumber\\ 
&&\times f \frac{ \langle m_{\rm cl} \rangle}{m_\star} \sigma^{3/2}~,
\end{eqnarray} 
where $m_\star$ is the mass of the field stars, $\langle m_{\rm cl} \rangle$ is the average globular cluster   mass,
and $f$ is the initial number  fraction of globular clusters~(after galaxy formation) to the total number of stars in the galaxy.
Since  the term in square brackets in equation~(\ref{eq5}) is a constant,  the globular cluster  merger model predicts $M_{\rm NSC} \propto \sigma^{3/2}$.

In the MW, the initial  total number  of  globular clusters, $N_{\rm cl}(0)$, can be recovered based on 
 their observed luminosity function and semi-analytical modeling 
 of mass-loss due to stellar evolution and due to  tidal interaction with the Galactic environment.  
\citet{Diederik09} derived a survival fraction of $0.004$ for a minimum initial cluster mass $M_{\rm min}=5000~M_{\odot}$. 
Assuming 100 present-day globular clusters we have that the total initial number of clusters in the Galaxy is
 $10^{4.4}$.  
 Considering only those GCs that are associated with the bulge on a Hubble time, this gives $N_{\rm cl}(0)=10^{3.9}$, which
compared to the total stellar  mass of the Bulge,
$\sim10^{10}~M_{\odot}$,  yields  $f\approx10^{-6}$.
We  can rewrite equation~(\ref{eq5}) as
\begin{eqnarray}\label{eq6}
M_{\rm NSC}&=& 3\times 10^{7} M_{\odot} \left( \frac{f}{10^{-6}}   \right) \left( \frac{\ln \Lambda}{3}   \right)\left( \frac{m_\star}{M_{\odot}}\right)\left(\frac{\langle m_{\rm cl} \rangle}{10^5 M_{\odot}}\right)^{3/2} \nonumber \\
&& \left(\frac{t}{10^{10} {\rm yr} }\right)^{1/2} \left(\frac{\sigma}{50 {\rm km~s^{-1}}}\right)^{3/2}~.
\end{eqnarray}
Despite its simplicity, our model reproduces   the observed $M_{\rm NSC} - \sigma$ relation both in slope and normalization.
 
 \subsubsection{Dissipative model}
 \citet{MKN:06}  proposed a \NSC \ in-situ star formation model  regulated by momentum feedback.
This model  is  an extension  of the argument proposed by \citet{King} to explain the $M_{\bullet}-\sigma$ relation, and
 invokes the formation of a massive nuclear cluster due to  gas inflow  and accumulation at the center of the galaxy 
 during the early phases of galaxy evolution. Stellar winds and supernovae from a young nuclear cluster with a standard IMF produce 
 an outflow with momentum flux given by 
 \begin{equation}
\dot{\Pi}\approx \lambda L_{\rm Edd}/c=\frac{ \lambda4\pi G M_{\rm NSC}}{k},
\end{equation}
 where $L_{\rm Edd}$ is the Eddington luminosity calculated by the stellar mass, 
 $c$ is the speed of light, $\lambda\sim 0.05$ is related to the mass fraction of massive stars,
and $\kappa\equiv 0.398{\rm cm^2~g^{-1}}$ is the electron scattering opacity. 
 The outflow is initially momentum conserving and produces an outward force on the gas in the bulge,
 whose weight is $W(r)=GM_{\rm gas}(r) M(r)/r^2$, with $M_{\rm gas}(r)$ 
the enclosed gas mass and $M(r)$
 the total enclosed mass of the galaxy. For an isothermal potential one finds
  \begin{equation}
W=\frac{4f_g}{G}\sigma^4~,
\end{equation}
where  $f_g=0.16$ is the baryonic mass fraction~\citep{Spergel}. 
 Requiring that the momentum output produced by the nuclear cluster balances the weight of
 the gas leads to the relation
 \begin{eqnarray}\label{eq7}
 M_{\rm NSC}&=& \frac{f_g\kappa}{\lambda\pi G^2}\sigma^4~= \\
 && 3\times10^{7} M_{\odot}\left(\frac{f_g}{0.16}\right)\left(\frac{\lambda}{0.05}\right)^{-1} \left(\frac{\sigma}{50{\rm km~s^{-1}}}\right)^4~.\nonumber
 \end{eqnarray}
A similar scaling relation can be derived in protogalaxies with  non-isothermal dark matter halos~\citep{MM12}.
 
 This model is very attractive because it contains no free parameters. 
 In addition, equation~(\ref{eq7}) is in good agreement with the $M_{\rm NSC} - \sigma$ correlation  reported by \citet{F:06}: 
  \begin{eqnarray}
\log  \left(\frac{M_{\rm NSC}}{M_{\odot}} \right)&=&(6.91\pm0.11)  \\
&+&(4.27\pm0.61)\log\left( \frac{\sigma}{54~{\rm km~s^{-1}}}\right)~,\nonumber
  \end{eqnarray}
but, as previously mentioned, it is at odds with  \citet{G12} who finds  \NSC\ scaling relations considerably 
shallower than the corresponding \sbh\ scaling relations. The difference between these two studies  
was due to the proper exclusion of nuclear disks in the sample of  \citet{G12}  and  
 the larger sample of \NSC s  used in this latter work. For these reasons we consider the results of
Graham more robust and conclude that
the   McLaughlin et al.  model provides a poor description of the observed $M_{\rm NSC} - \sigma$ correlation.
 This suggests  that momentum feedback may be not relevant, which
would be expected if the \NSC s  originated
 elsewhere  and were subsequently deposited into
their host galaxy centers.

\subsection{$R-M_{\rm NSC}$ relation}
The size of galactic nuclei clearly correlates with their luminosity, in the sense that 
brighter \NSC s have  larger effective radii.
The relation is approximately~\citep{Cote}
\begin{equation}
R\propto \sqrt{\mathcal{L}_{\rm NSC}}~,
\end{equation}
where $R$ is the \NSC \ effective radius (or half-mass radius) and $L$ its total luminosity.
This relation is consistent  with the extrapolation of the $R_{eff}- \mathcal{L}$ relation for elliptical galaxies.

\subsubsection{Dissipationless model}
In the merger model the radius of the nucleus increases with increasing total mass as globular clusters
merge.  The brighter, larger mass nuclei are therefore predicted to be spatially very extended.
\citet{AM12-2} used simple energy arguments to derive the  size-mass relation for \NSC s in
galaxies with no central \sbh . 
For the sake of completeness we repeat this simple calculation in what follows.
After a merger the \NSC\ energy, $E_f=-G M_{f}^2/2R_f$, equals the energy of the nucleus before the merger, $E_i$,
plus the energy brought in by the globular cluster: 
\begin{equation}
E_f=E_i+E_o+E_b~,
\end{equation}
with  $E_b$ the internal binding energy of the cluster,  and  $E_o\approx -G m_{\rm cl} M_{i}/2R_{i}$ its orbital energy before the merger.

The equations above permit expressing the mass, energy, and radius of the
nucleus recursively as
\begin{eqnarray}
M_{j+1}=(j+1)M_{1},~~~~~~~~~~~~~~~~~~~~~~ \label{rec1}\\
jE_{j+1}=\left( j+1 \right)E_{j}+jE_1,~~~~~~~~~~~~~~~~~~~\label{rec2}\\
\left( j+1\right)^2 R^{-1}_{j+1}=j\left( j+1\right)R^{-1}_j +R^{-1}_1, j=1,2,3,... ~~~~~~~~\label{rec3}
\end{eqnarray}
where the subscript 1 denotes the initial nucleus, and, by assumption, $M_1=m$.
Equations~(\ref{rec1}-\ref{rec3}) imply $R\propto M^{0.5}$ at the time the mass of the nucleus is still comparable to
the mass of the infalling clusters.
After many mergers the nucleus is much more massive than one globular cluster and the relation steepens to $R\propto M$. After
25 mergers, equations~(\ref{rec1}-\ref{rec3})  imply $R\approx 5 R_1$. The
 half-mass radii of globular clusters are $\sim3~$pc \citep{jord}, irrespective of their luminosity, so for a nucleus assembled from 25
mergers, $R\sim 15$ pc. Such a value  is in reasonable agreement with the measured half-mass radii  for the brightest nuclei. 
We conclude that a model that attributes the origin of \NSC s to the mergers of globular clusters at 
the centers of galaxies is consistent with the sizes and luminosities of the nuclei.

In a number of galaxies   \NSC s are observed to coexist with \sbh s. An example of such systems  is the MW
for which the mass of the \NSC, $M_{\rm NSC}\sim 10^7~M_{\odot}$~ \citep{LZM,S08}, is somewhat comparable to the mass of the 
central black hole, $M_{\bullet}\sim 4\times 10^6~M_{\odot}$~\citep{Ghez08,Gil}. A \sbh\ will disrupt clusters  that
pass within the radius
\begin{equation}
r_\mathrm{disr} \approx 2
\left(\frac{\sigma_\mathrm{NSC}}{5\sigma_K}\right)^{2/3}
\left(\frac{r_\mathrm{infl}}{r_K}\right)^{1/3} r_K~,
\end{equation}
where $\sigma_K$ is the one-dimensional velocity dispersion of a globular cluster, $r_K$ its core radius, 
$\sigma_{\rm NSC}$  is the velocity dispersion in the \NSC\ and $r_{\rm infl}=G M_{\bullet}/\sigma_{\rm NSC}^2$ is the influence radius of the \sbh. 
In the presence of a \sbh, the dependence of the half-mass radius $R$ on $M_{\rm NSC}$  is substantially weaker
than the observed correlation,  
$R\sim M_{\rm NSC}^{0.2}$~\citep{AM12-2}, due to the fact that the size of the
\NSC\ is determined by the fixed tidal field from the \sbh.  
When a stellar cluster is disrupted, stars that were initially within the cluster core will 
 redistribute  locally at a distance $r_\mathrm{disr}$ from  the \sbh . The 
density profile of the \NSC\ will therefore have a core of characteristic size $\sim r_{\rm disr}$.

 \subsubsection{Dissipative model}
 Although the gas model remains somewhat more qualitative there are some indications that 
  the observed  $R-M_{\rm NSC}$ relation might be difficult to reconcile with a purely 
 dissipative scenario.
 
\citet{B07} performed fully self-consistent  chemodynamical simulations to 
 investigate how \NSC s can form  through dissipative gas dynamics. He found that compact nuclei
 can be formed via dissipative, repeated merger of gaseous (or stellar) clumps that develop from 
nuclear gaseous spiral arms due to local gravitational instability.  These computations
showed that  fainter \NSC s are likely to have a more diffuse configuration
due to more negative feedback from SNe~II which in turn can prevent \NSC s from forming in faint galaxies.
The fact that \NSC \ formation is more-strongly suppressed by stronger feedback effects in
less-luminous galaxies would explain why brighter dwarf galaxies (dE) are more likely to contain \NSC s~\citep{VdB}.
However, simulations also show that more massive \NSC s are  less extended than their lower mass counterparts, 
which is   the opposite of the observed trend~\citep[see Figure~15~of][]{B07}.

Further investigation of the above inconsistency is needed in order to determine
how the results depend on the resolution limit of the simulations  and on the adopted initial conditions.

\section{Formation of \NSC s }  
\subsection{Phase-space constraints}
The above discussion gives some  level of 
reliability to the hypothesis of a dissipationless, merging formation for \NSC s.
 We now wish to  test whether  the merger of globular clusters can result in a \NSC\ similar to that of the MW.
   As a first step one can derive the globular cluster parameters required to give a peak density, $\rho_{\rm NSC}$, equal to the observed 
density of the  \NSC\ in the MW, and see whether such parameters 
are reasonable when compared to  typical globular cluster properties \footnote{I am indebted
to D.~Merritt for suggesting this calculation.}.

The stellar cluster central phase space density is given by
\begin{equation}
\frac{\rho_k}{\sigma_K^3}=\frac{9}{4\pi G}\frac{1}{\sigma_Kr_K^2}~,
 \end{equation}
 where $\rho_K$ is the cluster core density.
 After inspiral, we require $\rho_{\rm NSC}/\sigma_{\rm NSC}^3\lesssim {\rho_k}/{\sigma_K^3}$, or
 \begin{equation}
\rho_{\rm NSC}\lesssim \frac{9}{4\pi G}\frac{\sigma_{\rm NSC}^3}{\sigma_Kr_K^2}~,
 \end{equation}
so that
  \begin{eqnarray}
\rho_{\rm NSC}&\lesssim& 1.5\times 10^7 M_{\odot} {\rm pc^{-3}}
 \left( \frac{\sigma_{\rm NSC}}{100 {\rm km~s^{-1}}} \right)^3 
  \nonumber \\
&&  \times \left( \frac{\sigma_{K}}{10 {\rm km~s^{-1}}} \right)^{-1}\left( \frac{r_K}{{\rm 1pc}}\right)^{-2}~.~\label{e1}
 \end{eqnarray}
 
The observed \NSC\ density within $0.5~$pc at the Galactic Center~(GC) is
$\rho_{\rm NSC} \approx 10^6 M_{\odot} {\rm pc^{-3}}$~\citep[][]{M10}.
 Comparing this with equation~(\ref{e1}),  we obtain
 \begin{equation}
 \left( \frac{\sigma_{K}}{10 {\rm km~s^{-1}}} \right)\left( \frac{r_K}{{\rm 1pc}}\right)^{2}\lesssim 15~.
 \end{equation}
 
This crude calculation shows that for a  MW-like galaxy, 
the globular cluster parameters required to give the observed  peak density of the \NSC\ are quite reasonable.

\begin{figure*}
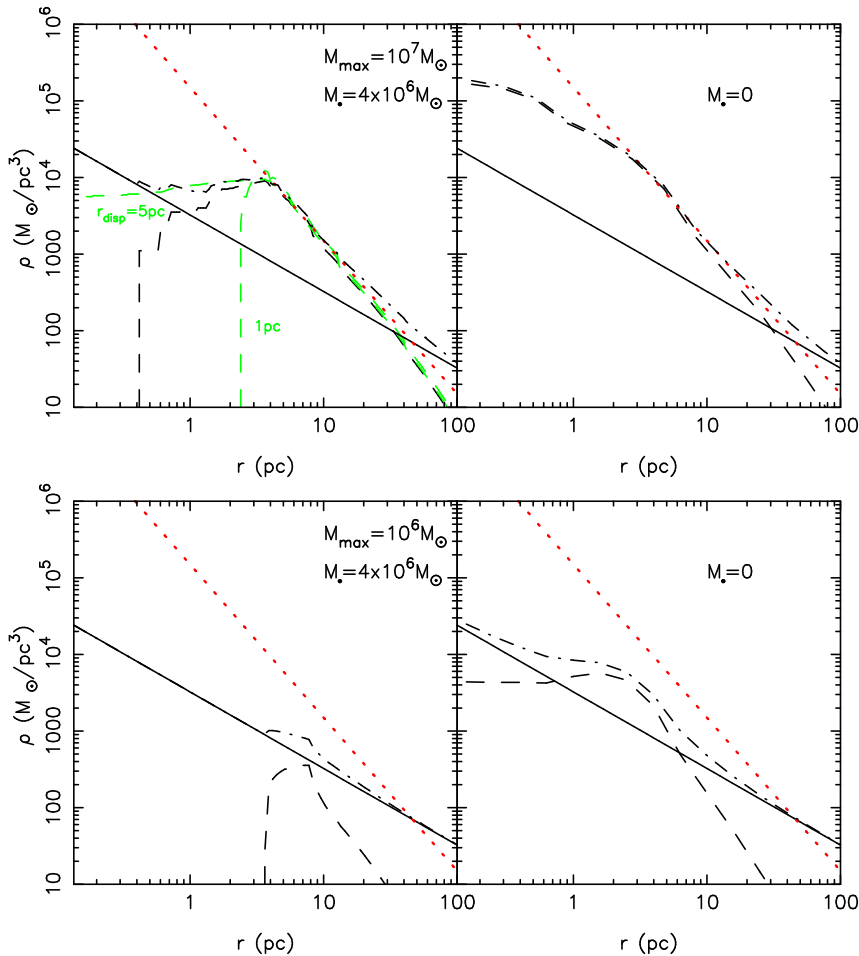

\centering
\includegraphics[width=.35\textwidth,angle=270]{Figure1a.eps}
\includegraphics[width=.35\textwidth,angle=270]{Figure1b.eps}
\caption{Formation of a \NSC\ in  a galactic bulge via cluster migration.
 Dashed and solid lines give respectively the density profiles 
 of the    \NSC\ after $10^{10}~$yr and the density of the
 background galaxy.  
Dot-dashed lines are the sum of galaxy and \NSC\ density profiles.
Red-dotted lines give the power-law
density model $\rho(r)=1.5\times10^5M_{\odot}\left(r/1~{\rm pc} \right)^{-2}$, 
representative of the observed radial distribution of stars in the MW \NSC~\citep{S09}.
We show results assuming that the total mass in stellar clusters is initially 
$10$ per cent of the total galaxy mass and for $M_{\bullet}=4\times10^6~M_{\odot}$(left panels), 
and $M_{\bullet}=0$~(right panels). The high mass truncation of the CIMF is $M_{\rm max}=10^7~M_{\odot}$ in the upper panels and 
$M_{\rm max}=10^6~M_{\odot}$ in the lower panels. Both large values of $M_{\bullet}$ and small values of $M_{\rm max}$ tend to reduce the excess density due to cluster infalls. 
In making these plots we have assumed 
the core mass of disrupted clusters is re-distributed after disruption over a region of finite extent  $r_{\rm disp}=3$pc.
The green  lines in the upper-left panel show results for
 $r_{\rm disp}=1$ and $5$pc. 
This figure clearly demonstrates how \sbh s can control the structure  of \NSC s forming through 
globular cluster merging. }\label{Fig1}
\end{figure*}

\subsection{\NSC\ formation  via cluster migration}
A simple analytical model 
is developed in what follows to 
evolve a population of stellar  clusters  subject 
to migration and dissolution in a galactic bulge and to calculate the influx of migrating clusters
into the center of the galaxy.
In this way we can  address the possibility that a substantial fraction of the  \NSC\ mass
in galaxies like the MW could have been assembled through cluster migration and mergers.

We assume that  the central properties of a stellar cluster~(i.e. $\sigma_K$ and $r_K$)  remain 
unchanged during inspiral and that $r_t>~r_k$, where $r_t$ is the cluster tidal~(limiting) radius given by~\citep{K62}
\begin{equation}\label{rtrking}
r_t= \alpha \frac{\sigma_K}{\sqrt{2}}\left(\frac{3}{r}\frac{ {d} \phi}{{d} r} -4\pi G \rho  \right)^{-1/2}~,
\end {equation}
with $\phi$ the galactic potential, and $\alpha$ a ``form factor'' that depends on the density distribution  within the cluster.
  For a King model with a large central concentration,   $\alpha\approx 1$. 
Equation~(\ref{rtrking}) includes the tidal force due to the radial gradient in the galaxy potential, the centrifugal force from the cluster's orbit
and assumes that the density of the cluster goes to zero roughly 
at the radius where the force acting to remove a star is balanced by the attracting force from the rest of the cluster. 

The tidal radius of a King  model orbiting in a galaxy with a power-law density profile, $\rho(r)=\rho_0\left(  r/r_0\right)^{-\gamma}$, and
containing  a \sbh\ at its center, is 
\begin{equation}\label{rtr2}
r_t\approx\frac{\sigma_K}{\sqrt{2}}\left[4\pi G \rho_0 \left(\frac{r}{r_0} \right)^{-\gamma} \frac{\gamma}{3-\gamma} +\frac{3GM_\bullet}{r^3}\right]^{-1/2}~.
\end {equation}
The radius at which the tidal force from the \sbh\ starts to dominate the tidal force
  from the stellar cusp (in galaxies containing both) is 
 \begin{equation}
r=\left(\frac{3(3-\gamma)M_{\bullet}}{\gamma 4\pi \rho_0 r_0^{\gamma}}\right)^{\frac{1}{3-\gamma}}~.
 \end{equation}
A King model also satisfies the relation:
\begin{equation}\label{mt}
Gm_t\approx \frac{\sigma_K^2r_t}{2}~,
\end{equation} 
where $m_t$ is the truncated mass of the globular cluster whose radius  is limited by the external tidal field.

Given the cluster central velocity dispersion,  equations~(\ref{rtr2}) and (\ref{mt}) can be combined  
 to  evaluate,  at any radius, the  cluster mass permitted by the galaxy tidal field
and  the mass dispersed along the orbit. This corresponds to a density enhancement with respect to the galactic background of
\begin{equation}\label{deltarho}
\Delta \rho(r)=\frac{ \sigma_K^2}{8\pi G r^2}\frac{dr_t}{dr}~.
\end{equation}
The resulting radial dependence of the density profile of stars in the growing \NSC\ is easily found to be
\begin{equation}
\Delta\rho(r) \propto \sigma_K^3\times\left\{ \begin{array}{ll}
	\sqrt{\gamma(3-\gamma)}~r^{\frac{\gamma-6}{2}} & \mbox{cusp}~, \\
	 r^{-\frac{3}{2}}/\sqrt{M_{\bullet}} & \mbox{black hole}~.
	\end{array}
	\right.
\end{equation}
Steeper density cusps  in the distribution of background stars give lower values of the density 
slope for the resulting \NSC. Larger values of $M_{\bullet}$ also give smaller densities near the center since
the cluster models start being truncated at larger radii and their mass is dispersed over larger spatial scales.

\subsubsection{Dynamical Friction}
We obtain the cluster orbits  by using the standard Chandrasekhar's
  dynamical friction formula~\citep{CH:43} for a self-gravitating cusp~\citep[e.g.,][]{MP04,J:10}:
  \begin{eqnarray}\label{eq:drdt}
{dr\over dt} &=& -2{(3-\gamma)^{3/2}\over 4-\gamma}\sqrt{G\over r_0} \\
\times &&\left({r\over r_0}\right)^{\gamma/2-2}\frac{F(\gamma)\ln\Lambda}{\sqrt{4\pi\rho_0 r_0^3}}{m_{\rm cl}} ~. \nonumber 
  \end{eqnarray}
  The coefficient $F$  is a  function of $\gamma$, with  $F=(0.193,~0.302,~0.427)$ for
$ \gamma=(1,~1.5,~2)$.
We set an initial  limiting  radius of $40~$pc.
 If $r_t > 40~$pc (i.e. the model is not truncated),  
   integrating  equation~(\ref{eq:drdt}) yields
\begin{eqnarray}\label{sfc1}
r(t)&=&\Big[ r_{in}^{\frac{6-\gamma}{2}} - \frac{(3-\gamma)^{3/2}(6-\gamma)}{(4-\gamma)} \times\\
&& \sqrt{\frac{G}{4\pi\rho_0}}r_0^{-\gamma/2}F(\gamma) \ln\Lambda~  m_{\rm cl}\times t \nonumber
\Big]^{\frac{2}{6-\gamma}}~,~~~~
\end{eqnarray}
which gives a timescale to reach the center of the galaxy:
\begin{eqnarray}
\tau_{\star}
={10^{10}~{\rm yr}}\frac{(4-\gamma)}{(3-\gamma)^{3/2}(6-\gamma)F(\gamma)}  \\
\ln \Lambda_{3} ^{-1}\sqrt{ \rho_{0,5}} r_{0,700}^3 m_{{\rm cl},6}^{-1}   \left(\frac{r_{in} }{r_0}\right)^{\frac{6-\gamma}{2}},
\nonumber
\end{eqnarray}
with  $\rho_{0,5}=\rho_0/5~M_{\odot}{\rm pc^{-3}}$, $r_{0,700}=r_0/700~{\rm pc}$, 
$\ln \Lambda_3=\ln \Lambda/3$
and $m_{{\rm cl},6}=m_{\rm cl}/10^6~M_{\odot}$. 
The coefficient depending on $\gamma$ is approximately  $1$.

After  a cluster starts being truncated by the galactic tidal field, its mass also becomes  
a function of radius. 
In this case, we computed the cluster orbit  by setting $m_{\rm cl}=m_t$ in
equation~(\ref{eq:drdt}), obtaining
\begin{eqnarray}\label{sfc2}
r(t)=\left[ r_{in}^{3-\gamma}-\frac{(3-\gamma)^3}{(4-\gamma)\sqrt{2\gamma}}\frac{ 
r_0^{-\gamma}}{4 \pi G \rho_0 }F(\gamma)\ln \Lambda\sigma_k^3 \times t \right]^{\frac{1}{3-\gamma}}~.~~~~
\end{eqnarray}
This equation  takes into account
 mass loss due to the interaction with the galactic tidal field, but
ignores the possible presence of a \sbh~\footnote{These equations
also assume  
circular orbits. This is consistent with the well known effect of orbital 
circularization due to dynamical friction~\citep[e.g.,][]{CPV,Hashimoto}.} .
Equation~(\ref{sfc2}) implies that the massive object comes to rest at the center of the
stellar system in a time
\begin{eqnarray} \label{Tdf-sfc2}
\tau_{\star}&=&
3\times 10^{10}{\rm yr}  \frac{(4-\gamma)\sqrt{\gamma}}{(3-\gamma)^3F(\gamma)} \\
&&\ln \Lambda_{3} ^{-1} \rho_{0,5} r_{0,700}^3\sigma_{K,10}^{-3}  \left(\frac{r_{in}}{r_{0}}\right)^{3-\gamma}~, \nonumber
\end{eqnarray}
with  $\sigma_{K,10}=\sigma_K/10~{\rm km~s^{-1}}$ and
the coefficient depending on $\gamma$ equals to $(2.8,~4.27,~9.4)$ for $ \gamma=(1,~1.5,~2)$.

These basic formalism  assumes that mass loss from the clusters is entirely due to their  interaction 
with the external tidal field and it omits mass loss~(evaporation) due to internal dynamics.
Since the only clusters that can contribute to the growth of a  \NSC\   are very massive systems with relaxation times
longer than their dynamical friction timescale this simplification is quite reasonable and 
does not alter our results in any important way.
However, dynamical evaporation due to internal dynamics is also accounted for, 
 in the sense that we do not follow the evolution of clusters which are  disrupted
(evaporated) on timescales  shorter than the relevant dynamical friction timescales.

\begin{figure*}
\centering
\includegraphics[width=.3\textwidth,angle=270]{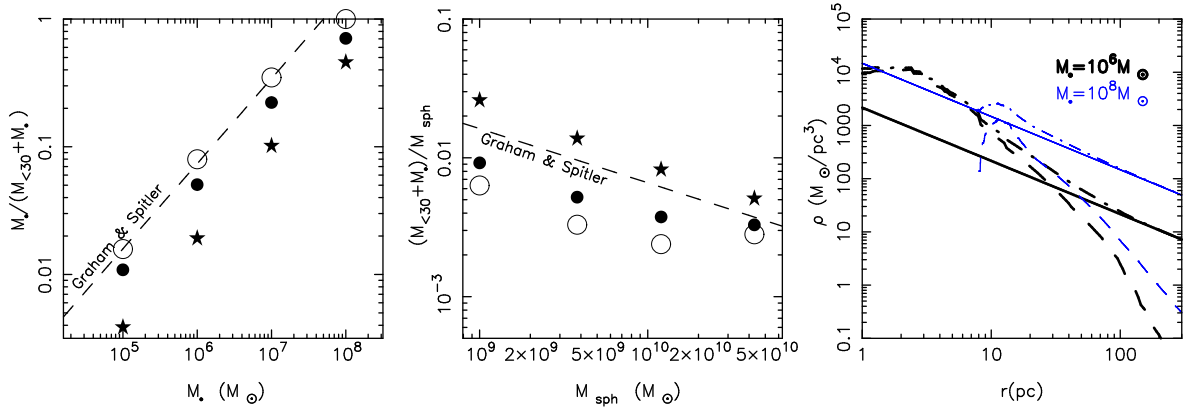}
\caption{
Effect of a galactic center  \sbh\ on the \NSC\  properties.
We assume $M_{\rm max}=10^7~M_{\odot}$ and
that the total mass in clusters is initially 
$0.05\times M_{\rm sph}$~(open circles), $0.1\times M_{\rm sph}$~(filled circles), or $0.3\times M_{\rm sph}$~(star symbols). 
The properties of the galaxy are changed with $M_{\bullet}$
according to the observed scaling relations between \sbh s and the properties of their
host spheroid. 
In the left panel we plot the ratio $M_{\bullet}/(M_{\bullet}+M_{<30})$ as a function
of $M_{\bullet}$, where $M_{<30}$ is the mass deposited 
in the inner $30~$pc of the galaxy. Dashed line is the observed correlation: equation~(1) in
\citet{GS:09}. This plot measures the relative importance of $M_{\bullet}$ to the
total mass in the nuclear components for different black hole masses.
The middle panel gives the importance (in terms of mass) of the nuclear component~(\sbh\ + \NSC )
relative to the host spheroid mass. Dashed line is the observed correlation given by equation~(2)
in \citet{GS:09}.
As an example, the right panel displays the  density profile of background galaxy~(solid lines),
\NSC\ (dashed lines)  and the sum of them~(dot-dashed lines) 
for an initial  total mass in clusters equal to $0.1\times M_{\rm sph}$
and for $M_{\bullet}=10^6$ and $10^8~M_{\odot}$.
Clearly,   larger \sbh s correspond to
 \NSC s with lower central densities. For $M_{\bullet}\gtrsim 10^8$, the \NSC \   densities 
 remain below the density of the galaxy at all radii.}\label{Fig2} 
\end{figure*}

\subsubsection{Formation of the MW \NSC\ } \label{MW}
Luminosity profiles of galaxies are well approximated by power laws  with  $1<\gamma<2$ at radii smaller
than the effective radius of the stellar spheroid~\citep[$R_{eff}$;][]{TG05}. Since for  MW-like galaxies 
the only clusters that can reach the GC in one Hubble time are those initially 
at $r \lesssim R_{eff}$~\citep{Milos04}, and that within these radii the 
baryonic matter dominates the galactic potential, 
we represent the galaxy bulge by using a simple power-law density model:
$\rho(r)=\rho_0\left({r}/{r_0}\right)^{-\gamma}$,  with $\rho_0=(3-\gamma)M_{\rm sph}/4\pi r_0^3$~
\footnote{This expression assumes that the density follows a \citet{Dh93} 
profile at large radii, i.e., $\rho(r)\sim r^{-4}$ for $r\gg r_0$.},
$M_{\rm sph}=10^{10}~M_{\odot}$,  $r_0=700~$pc and $\gamma=1$~\citep[e.g.,][]{MD07}. 
The scale length $r_0$, is  related to the bulge effective radius  via~$R_{eff}/r_0=(1.8,1.5,1)$ for $\gamma=(1,1.5,2)$.
We assume that the clusters have  initially the same  distribution of stars in the galaxy~\citep[e.g.,][]{AM:11} and we assign their masses
using the cluster initial mass function~(CIMF), ${\rm d}n/{\rm d}M\propto M^{-2}$~\citep{bik,degrijs}, and limiting mass values of
$M_{\rm min}=10^2~M_{\odot}$; $M_{\rm max}=10^6-10^7~M_{\odot}$.

%The vast majority  of stars in a galaxy form in 
%bound stellar clusters~($\sim 5~\%$) of these  systems
% survives the gas embedded phase~\citep{tt78,Lada+84}.
The fraction of all star formation that occurs in gravitationally bound stellar clusters is usually referred to as
 ``cluster formation efficiency''. 
The cluster formation efficiency in the MW is $\sim 10~\%$~\citep{LL03}.
Accordingly,  in our model we assume that the total mass in clusters 
is initially $M_{\rm cl}=0.1\times M_{\rm sph}$.

We start by   identifying   a ``dissolution time" for globular clusters due to dynamical evaporation 
 as a function of their initial mass~\citep{BM:03,GB:08,Lamers+10}:
 \begin{equation}\label{tdiss}
t_{\rm dis}=t_0\left( \frac{m_{\rm cl}}{M_{\odot}} \right)^{\beta} 
\end{equation}
with $\beta=0.7$, $m_{\rm cl}=m_t$ if the cluster is truncated, $t_0=0.3\times |d \Omega/d\ln r|^{-1}$ and $\Omega$ 
the angular velocity at the galactocentric radius $r$.
We then remove  clusters with total lifetime, $t_{\rm dis}/\beta$, 
shorter than the dynamical friction time-scale, $\tau_{\star}$. 
This procedure
 reduces the initial total mass in globular clusters from $10^9~M_{\odot}$ to 
 $\sim 5\times 10^{7}~M_{\odot}$ for $M_{\rm max}=10^6~M_{\odot}$, and to
$\sim 2\times 10^8~M_{\odot}$ for $M_{\rm max}=10^7~M_{\odot}$.
By comparing  equation~(\ref{Tdf-sfc2}) with equation~(\ref{tdiss})   one can easily show that 
if   $\gamma \lesssim 3/2$ and $t_{\rm dis}/\beta > \tau_{\star}$ initially, this latter
  condition remains satisfied during inspiral.
  
We set the core radius of the globular clusters to be $r_K=1~$pc, roughly 
equal to the median value of  the core radii listed in the 
Harris's compilation~\citep{Harris} of Galactic globular clusters.
If a cluster reached its tidal disruption radius
its  core mass  is redistributed  uniformly at the radius of disruption over a region of extent
$r_{\rm disp}=3~$pc. But we  note that our results do not strongly depend 
 on the particular value for this parameter, which only affects the NSC density profile at very 
small radii~($\lesssim 3~$pc). Within these central regions we  expect that
other dynamical processes~(e.g.,  two-body relaxation, mass segregation) will modify 
 the NSC  density profile with respect to the simple predictions 
of our Monte-Carlo experiments~(we discuss this point in more detail below).
When a cluster enters the inner $1~$pc,  its mass was redistributed using a Plummer
sphere model of total mass~$m_t(\rm 1pc)$ and core radius $r_K$.

In addition, we make the reasonable assumption that the distribution of field stars is
constant when computing the relevant dynamical
friction time-scales. Given  that the  total mass in clusters 
is  only $10\%$ of the bulge mass initially,
the density profile of the galaxy at intermediate and
large radii~\citep[$\gtrsim30~$pc;~e.g., see Figure~4 of][]{AM:11} and consequently 
the dynamical friction timescales
are not significantly affected by such 
simplification.
However, to approximately account for the 
mass that  is progressively accumulated into  the nucleus,  the orbital decay of clusters 
was computed in  order of increasing migration time; then, for any inspiral,  
 we added the quantity  $3GM_{\rm acc}(<r)/r^3$
to the terms in square brackets in equation~(\ref{rtr2}),
with $M_{\rm acc}(<r)$  the accumulated mass within $r$ due to 
clusters  with shorter orbital decay times.

 Finally, we computed the  \NSC\ density  at any
 radius by summing up the contribution, $\Delta \rho(r)$,
 of all clusters that within $10^{10}~$yr reached that radius.
 After this time the total mass left in stellar clusters  is
$\sim  10^7~M_{\odot}$,
almost independent  on the initial value of $M_{\rm max}$.
The initial mass of the globular cluster system is therefore reduced 
from $10~\%$ to 
 $0.1~\%$ of the total  Bulge mass after one Hubble time
 due to dynamical dissolution of the less massive systems, and 
  also due to  massive clusters that inspiral into center of the galaxy and dissolve locally  
  to form the stellar nucleus.
Figure~\ref{Fig1} gives the results of such calculations. 

In agreement with \citet{AM:11} we found that the mass of the forming \NSC\ 
is mostly ($\gtrsim 90\%$) composed of clusters 
with initial  masses $\gtrsim 0.1\times M_{\rm max}$, suggesting that the low mass clusters scarcely contribute 
to  \NSC\ formation in galaxies. Instead, our results  very much depend  on the value of $M_{\rm max}$, 
since only very massive clusters can arrive in the central regions of the galaxy in a reasonable time
without being destroyed by the galactic tidal forces in the process.
 
In galaxy models without a \sbh, \NSC\ formation is more efficient, and 
even for $M_{\rm max}=10^6~M_{\odot}$ a \NSC\ forms  in the central few parsecs of the galaxy.
If a \sbh\ is present  at the center of the galaxy, star clusters that 
come closer than $r_{\rm disr}$ from the center are disrupted
and, as a result, the \NSC\ density  is  limited inside this radius~(left panels in the Figure~\ref{Fig1}).
 For a \sbh\ mass of $4\times 10^6~M_{\odot}$ and $M_{\rm max}=10^6~M_{\odot}$,
no clusters penetrate radii smaller than $\sim 7~$pc. From the lower-left panel in Figure~\ref{Fig1}
we can see that, in this case, the presence of a \sbh\ will strongly  inhibit the formation of a \NSC.
Taking $M_{\rm max}=10^7~M_{\odot}$~(upper-left panel) results in a smaller radius of the \NSC\ core 
and  higher central densities,  since the most massive clusters can reach smaller galactocentric radii~($\sim2~$pc).

A model  that starts with a high mass truncation of $M_{\rm max}=10^7~M_{\odot}$  reproduces 
 the  observed mass density profile of stars  in the Galactic \NSC~(red-dotted lines in Figure~\ref{Fig1})
outside  $3~$pc. We conclude that a scenario in which 
a large fraction of the mass of the Milky Way \NSC\
 is due to infalling globular 
clusters is  in good agreement  with current observational constraints.

  The results presented here should be interpreted  with caution when comparing  the detail of the 
\NSC \ density profile to observations. Our modeling  includes  a series of simplifications
that can have some impact on the inner structure of the resulting \NSC. 
For example, consideration of the internal dynamics and mass spectrum in the globular clusters could 
enable mass segregation which can increase $\sigma_K$ and then allow the cluster stars to reach 
much smaller radii.  A realistic treatment  of these effects will require careful $N$-body simulations and  is beyond 
the scope of this paper.
Nevertheless, we note that  the results of our computations that include 
a  central \sbh\ are very similar to those obtained in \citet{AM12-2} via $N$-body
 simulations. In both cases, the density profile 
 that results after the final inspiral event is characterized by a core of roughly the tidal disruption radius 
 of the most massive clusters, $\sim 3~$pc, and an envelope with density that falls off as $\rho\sim r^{-2}$.
 These properties are similar to those of 
the MW \NSC, with the exception of the core size, which in the MW is somewhat smaller~($\sim 0.5~$pc). 
\citet{M10} showed that such a core will shrink 
substantially via gravitational encounters in a time of 10 Gyr as the 
stellar distribution evolves toward a Bahcall-Wolf cusp~\citep{BW:76}. 
In our computations cluster inspiral occurs more or less 
continuously over the lifetime of the galaxy. The core resulting from the combined effects of 
cluster inspiral and relaxation would therefore be somewhat smaller than 
 that we found above, and closer to the observed 
core size~(we discuss this point in more details below in \S \ref{mbh}).
   
 \begin{figure}
~~~~~~~~~~~~~~\includegraphics[width=.8\textwidth,angle=270]{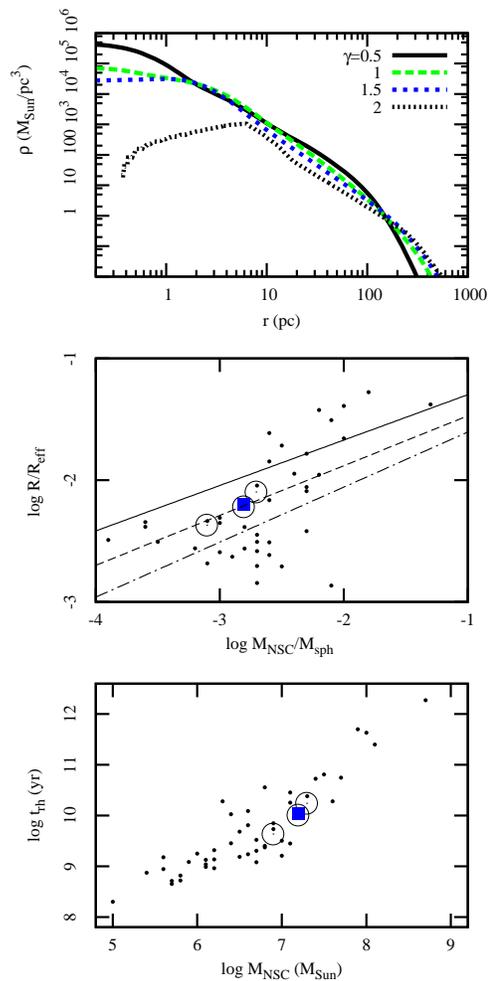}
\\ 
\caption{ Upper panel gives the density profile of 
\NSC s  after $10^{10}~$yr in galaxy models with inner density profile slope $\gamma=(0.5,1,1.5,2)$.
In these computations we set $M_{\rm sph}=10^{10}~M_{\odot}$, 
$R_{eff}={\rm 1~kpc}$, $M_{\bullet}=0$, and
$M_{\rm cl}=0.1\times M_{\rm sph}$. 
The high mass truncation of the CIMF is $M_{\rm max}=10^7~M_{\odot}$.
The   largest 
\NSC\ central  densities are found in models with small values of $\gamma$.
In the middle and lower panels the properties of the \NSC\ models are compared to those of real \NSC s in early type 
galaxies~\citep[dots are data from][]{Cote,seth}.  
The middle panel plots the nuclei in the size ratio versus mass ratio plane. 
Open circles correspond to  the \NSC\ models of the top panel with $\gamma=(0.5,~1,~1.5)$, where 
the relaxation time
and  effective radius of the \NSC\ 
decrease with increasing  $\gamma$. The blue-filled squares represent the model in the upper right panel
of  Figure~\ref{Fig1}. Lines indicate the critical value of $R/R_{eff}$ over which nuclei expand 
for galaxy models described by Einasto indices $n=2$~(continue line), $n=3$~(dashed line) and $n=4$~(dot-dashed line), 
\citep[for comparison see Figure~1c in][]{M09}.
The lower panel shows the dependence of  \NSC\ half-mass  relaxation time on
  its total mass. }\label{Fig3} 
\end{figure}

  \subsubsection{The role of MBHs}
Figure~\ref{Fig2}  shows the results of additional computations that explore
the effect that varying $M_{\bullet}$ has on  the structure of the  \NSC.
In these plots we set $M_{\rm max}=10^7~M_{\odot}$ and
we relate the mass of the galaxy to $M_{\bullet}$
through the 
relation~$\log \left[M_{\bullet}/M_{\odot} \right]
 \approx 
 8.4+1.9 \log \left[ M_{\rm sph}/7\times 10^{10}M_{\odot}\right]$~\citep{GR12}.
 The galaxy velocity dispersion,  $\sigma$, is derived from the $M_{\bullet}-\sigma$ relation~\citep{FM,G01},
while the model scale length, $r_0$, is 
obtained from the effective radius: 
$R_{eff}\approx 1~{\rm kpc}~(M_{\rm sph}/{10^{10}~M_{\odot}})/(\sigma/100~{\rm km~s^{-1}})^2$,
where here the normalization    reproduces approximately the observed 
effective radii of elliptical galaxies~\citep{FORBES,GW:08}.  
In the left panel of Figure~\ref{Fig2}
 we show the increasing dominance of the central \sbh\ over the 
 \NSC \ stars for more massive galaxies. Our results agree reasonably well with studies of galaxies
 at high mass end with $M_{\bullet}>10^8~M_{\odot}$, and which exclude the presence of stellar nuclei in such 
 systems~\citep[e.g.,][]{HR}. 
The middle panel of~Figure~\ref{Fig2} displays how the total  central mass in the nuclear component~(i.e., \sbh\ + \NSC )
divided by the stellar mass of the host spheroid varies with this latter quantity. Also in this case
our results agree reasonably well with the observed correlation~(dashed line in the plot).
 
 In order to reproduce the  \NSC\ masses  obtained from observations, 
 our model would require a high mass truncation 
 of the CIMF of  $M_{\rm max}=10^7~M_{\odot}$ and that $\lesssim 10~\%$ of stars in galaxies  originate in 
 stellar clusters.  
 Such a result is  in agreement  with the conclusion of \citet{Nathan} that 
the globular cluster  infall model strongly  under-predicts  the observed NSC masses 
 when assuming the present day number of globular clusters in galaxies.
This assumption  might  be incorrect however, given that
the number of globular clusters in a galaxy drops rapidly with time
due to dynamical disruptions~\citep[e.g.,][]{Vesperini97,Vesperini98,FZ:01,Gieles}.
In addition, the cluster formation efficiency and the cluster disruption rate
 also vary substantially with cosmic time,   peaking 
at early epochs, in gas-rich disk galaxies~\citep{Diederik1}. 
In fact, up to 30~per~cent of all stars in the Universe could
have been  formed in bound stellar clusters~\citep{Diederik2}.

 The right panel of  Figure~\ref{Fig2} gives the density profile of a \NSC \ that forms 
 around a \sbh\ of mass $M_{\bullet}=10^6-10^8~M_{\odot}$ assuming an initial mass in clusters
 $M_{\rm cl}=0.1\times M_{\rm sph}$. 
Smaller black hole masses correspond to smaller tidal disruption radii 
for the infalling globular clusters. As a consequence,
the  \NSC\  peak density becomes progressively larger and the \NSC\ core radius smaller as $M_{\bullet}$ decreases.
The mass delivered  in the inner region of the galaxy is also a strong function of $M_{\bullet}$.
For $M_{\bullet}\gtrsim 10^8~M_{\odot}$, very little mass~($\lesssim10^6~M_{\odot}$) can be deposited in the inner 
 $\sim 30~$pc of the galaxy. In this case
 the \NSC\ density profile remains below the 
 stellar density of the background galaxy.

   \bigskip

 To summarize the results presented in this section, 
 if we postulate that  \NSC s grow slowly through globular cluster migration and merging, then
in order to form a \NSC\  similar to that of the MW,
our model will require $M_{\rm max}\sim 10^7~M_{\odot}$ and that
 $\sim 10$~per cent of the  Bulge mass originated in stellar clusters.
We  showed that this fraction  reduces to roughly 
$0.1~\%$ of the total  Bulge mass
after $10^{10}~$yr due  to  dynamical evaporation of the less massive clusters, and also
because more  massive clusters inspiral toward the galactic center  
and dissolve due to their interaction with the  galactic tidal field.
 
 The presence of a  pre-existing \sbh\  at the center of the galaxy 
can have a strong impact on the inner structure of the forming nucleus.
 If a  \sbh\ of mass $M_{\bullet}\sim10^6~M_{\odot}$ 
 sits  at the center of the galaxy, massive globular  clusters 
are disrupted at radii of order $\sim3~$pc and the  \NSC\ density  
profile will have a core of roughly this radius. 
\sbh s less massive than about $10^6~M_\odot$ do not  influence very much
the structure of the stellar nucleus during its formation since the tidal disruption radius of massive GCs is in this case of the order (or less than)
the core radius of the most massive clusters, and in either models, with or without \sbh , the \NSC\ density profile will have
 have  a central core of typical  size $r_k$. In low mass spheroids  after a few Gyr  such a core will most likely relax to
 a Bahcall-Wolf cusp  or,  in the absence of \sbh , undergo core-collapse.
  The central peak  density of the resulting  \NSC\ progressively declines 
as $M_{\bullet}$ increases.
\NSC s forming  around \sbh s with $M_{\bullet}\gtrsim10^8~M_{\odot}$  have such low central densities that they
would be more difficult to observe  as distinct galactic components; this appears to be  in agreement with observations
which reveal a lack of \NSC s in stellar spheroids  brighter than about $10^{10.5}~L_{\odot}$.

   \section{Dependence on $\gamma$}
   In the above discussion we have considered Dehnen's models with 
   inner density profile  slope $d \log \rho / d \log r=-1$. 
   In Figure~\ref{Fig3} we relax this assumption 
   and explore the formation of \NSC s in galaxy models with different values of $\gamma$.
   In these computations we adopt $M_{\rm max}=10^7~M_{\odot}$ and $M_{\rm cl}=0.1\times M_{\rm sph}$.
   The structural parameters defining the galaxy model  are
 $M_{\rm sph}=10^{10}~M_{\odot}$, $R_{eff}={\rm 1~kpc}$, and $M_{\bullet}=0$. 
   
On the basis of equation~(\ref{deltarho}) we would expect 
the final~(after $10^{10}~$yr)  densities in the inner $\sim 100~$pc of the forming \NSC\ to
 be lower in galaxy models with steeper density profiles, 
provided that the dynamical friction  timescale of  massive stellar clusters remains shorter than
 the Hubble time. 
 The results illustrated  in the upper panel of Figure~\ref{Fig3}  agree with 
 this basic prediction, suggesting  that, independently on $\gamma$,
    the sinking timescale 
 of massive clusters in these models
 is  short enough that they can always reach the center of the galaxy and form a compact nucleus. 

The middle and bottom panels in Figure~\ref{Fig3}   compare the structural properties of our model \NSC s 
to the properties 
of \NSC s  in early-type galaxies that were found to be nucleated in \citet{Cote}.
For each of the \NSC\ models an
approximation of  its effective radius is obtained as the effective radius  of the best fitting 
S\'{e}rsic model to the \NSC\ projected density profile within a galactocentric radius of $30~$pc. Given $R$, the
total \NSC\ mass, $M_{\rm NSC}$, 
is then obtained as the total mass within a radius twice the effective radius.
 A useful reference time is the relaxation time computed at  $R$. Setting
${\rm ln} \Lambda=12$,  and ignoring the possible presence of a MBH, the half-mass relaxation time is~\citep{Spitzer}:
\begin{equation}\label{th}
t_{\rm rh}=1.75 \times 10^5\frac{ \left[ r_h(\rm pc)\right]^{3/2} N^{1/2}}{(m/M_{\odot})^{1/2}}{\rm yr}~,
 \end{equation}
 where $N$ is the total number of stars and $m$ is the  mass of
 a single star~($m=M_{\odot}$ in Figure~\ref{Fig3}).
From Figure~\ref{Fig3} it is evident  that 
our \NSC\ models have structural  properties that are similar to  those of real nuclei.
 
 In the absence of a central \sbh\, the dynamical evolution of a nuclear cluster
  is a competition   between core collapse, which causes densities to increase, and heat input from the surrounding 
galaxy, which  causes the cluster to expand and densities to decrease~\citep{K90,Q96,M09}. 
In a double-Plummer-law galaxy model, the maximum size of a \NSC\ of mass $M_{\rm NSC}/M_{\rm sph}=10^{-3}$ in order to resist expansion is  $R/R_{eff}\approx 0.02$~\citep{Q96}.
 This simple condition, when compared to the middle panel of Figure~\ref{Fig3},
  implies that the core-collapse time of our models are  always shorter that the time for the nucleus 
to absorb energy from the rest of the galaxy. 
However, when using more realistic Einasto profiles to describe the surrounding  galaxy  our 
\NSC\ models appear to be close to the transition region between systems that are in the prompt
core collapse phase and systems for which  the evolution is driven by heating from the galaxy.
Our models will therefore either expand or contract depending on the
``compactness'' of the surrounding galaxy.

 \section{Orbital decay in the core of a giant elliptical galaxy}
Stellar spheroids more luminous than $\sim 10^{10.5}~L_{\odot}$
often host   high mass black holes~($M_{\bullet}\gtrsim 10^9~M_{\odot}$) at their center,
while  they show no evidence for nucleation.  Rather, the density profile of such ``giant'' elliptical galaxies
 is observed to be flat inside the influence radius  of the \sbh~\citep{Kormendy87,Lauer:02}.
Two factors have been invoked to explain the observed lack  of \NSC s in such  systems:~
(i)  \NSC s and \sbh s grow in competition from the same gas reservoir~\citep{Escala07}. If a \sbh\ forms first
this  can prevent, through its feedback, a \NSC\ from growing 
 if the gas accretion rate is smaller than the Eddington rate~\citep{Nay09}.
 (ii)  Massive black hole binaries forming during the last galaxy-galaxy major merger  destroyed  their host \NSC s
 by ejecting stars from the inner galactic regions~\citep{BG:10}.
 
Observations seem to point against both scenarios.
Picture (i) seems to break down  for \NSC-dominated galaxies, given the extremely low accretion rates observed
in bulgeless   galaxies and that at least few of them also contain central \sbh s. 
The model  of \citet{BG:10}  is disfavored due to  the fact that
what sets \NSC\ disappearance does not seem to be galaxy morphology. Rather, there is a  limit to
the \sbh/\NSC\ mass ratio that fixes a sharp transition from galaxies with \NSC\ and \sbh-dominated galaxies~\citep{NW12}.

\begin{figure}
\centering
\includegraphics[width=.73\textwidth,angle=270]{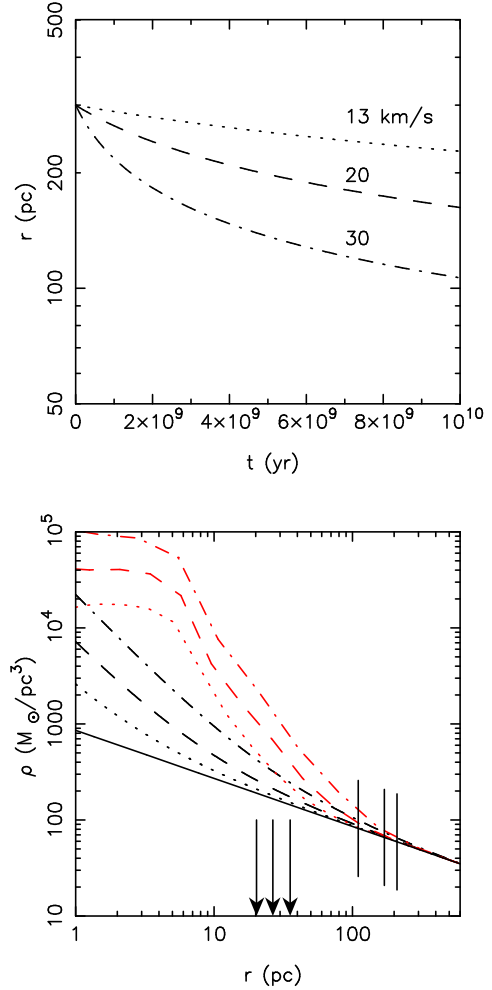}
\caption{Upper panel gives the orbital evolution of clusters with different values of the central velocity dispersion
in the core of a M87-like galaxy. The dynamical friction timescale for clusters in such models
is very long due to the lack of stars that move slower than the local circular velocity inside $r_{\rm infl}$.
Even after $10^{10}~$yr the cluster orbital  radii have hardly changed from their initial values.
The lower panel gives the density profile of the galaxy, assuming that the clusters did
 have enough time to migrate to the center: dotted, dashed  and dot-dashed curves  
show the results of the accumulation of  $500$ clusters with $\sigma_K=13;~20$ 
and~$30~{\rm km~s^{-1}}$ respectively. 
Red lines correspond to computations  without a \sbh\  at the center of the galaxy.
The galaxy background model is shown as a solid black line.
Vertical marks correspond to  the orbital radius of the cluster in the model with an 
\sbh\ after $10^{10}~$yr and starting from $r_{in}=300~$pc. Arrows give the tidal disruption radii of the clusters due to the \sbh .
At radii smaller than these our computations including a \sbh\ are 
no longer valid.}
\label{Fig4}
\end{figure}

Of course, it is possible that 
  nuclei are absent in bright galaxies 
because some mechanism prevents them  from forming   in the first place, or because 
they did not have time to reform  after they were destroyed by the scouring effect
of binary black holes.
In what follows, we show  that \NSC s might  in fact be difficult to form in such galaxies,
due to long dynamical friction time-scales of star clusters and also
 the little mass that the clusters can deliver  to the center due to the strong \sbh\ tidal field. 
In order to do so, we repeat a similar analysis to that presented in \S~\ref{MW}
but for a galaxy like M87, a giant elliptical with no evidence for \NSC.
We stress that the  problems discussed here and in \S~\ref{MW}
are substantially different. The influence radius of the \sbh\ in a galaxy like M87
extends much further out~(a few hundreds of parsecs from the center) than the tidal disruption radius 
of a massive globular cluster~($\sim 10~$pc), while for the MW we found $r_{\rm disr}\lesssim r_{\rm infl}$.
In the former case, the clusters will therefore move in a 
stellar background  whose velocity distribution is directly influenced
by the  presence of a  \sbh , which will strongly affect the timescale 
for cluster  inspiral as we now show.

The distribution of field-star velocities has the following form near a \sbh:
\begin{equation}
f(v_\star) =\frac{\Gamma(\gamma+1)}{\Gamma(\gamma-\frac12)}
\frac{1}{2^\gamma\pi^{3/2}v_c^{2\gamma}}
\left(2v_c^2 - v_\star^2\right)^{\gamma-3/2}~,  \\	
	\label{eq:fofv}
\end{equation}
where the normalizing constant
corresponds to unit total number.
This expression  gives the local
distribution of velocities at a radius where the circular velocity
is $v_c=(GM_\bullet/r)^{1/2}$, assuming that the density of field stars
follows ${\rho}(r)=\tilde{\rho}(r/\tilde{r})^{-\gamma}$.
The phase space density is zero for $v_\star\ge v_\mathrm{esc}=2^{1/2}v_c$.

As $\gamma \rightarrow 1/2$ the velocity distribution~(\ref{eq:fofv}) becomes progressively narrower, and for $\gamma=1/2$
all stars have zero energy; in other words,
the number of stars with $v_{\star}<v_c$ goes to zero as $\gamma$ approaches $1/2$. In the case
of a test particle moving in a circular orbit 
the standard Chandrasekhar's formula will therefore predict zero frictional force.
It turns out that  in this situation the  frictional force must  be computed using a more general formulation that also includes  the contribution
from stars moving faster than the test particle:
\begin{eqnarray} \label{dfa3}
\boldsymbol{f}_{\rm df}   &\approx& \boldsymbol{f}^{(v_\star<v)}_{\rm df}+ \boldsymbol{f}^{(v_\star>v)}_{\rm df}    = \\
&&-4\pi G^2  m_{\rm cl} \rho(r) \frac{\boldsymbol{v}}{v^3}\nonumber \times{\Big(}  {\rm ln \Lambda}\int_{0}^v dv_\star 4\pi f(v_\star) v_\star^2 \\
 &+&\int^{\sqrt{-2 \phi(r)}}_v dv_\star 4\pi f(v_\star) v_\star^2    \left[ {\rm ln} \left( \frac{v_\star+v}{v_\star-v} \right)-2\frac{v}{v_\star} \right]  \Big)~,~~~~~ \nonumber
\end{eqnarray} 
where  $ \boldsymbol{v}$ is the velocity of the infalling cluster. The
second term in parentheses represents the frictional force produced by stars moving faster than the massive particle.
$N-$body experiments verify the accuracy of this formula~\citep{AM12}.

The  fraction of stars that move slower than the local circular velocity 
can be computed as
\begin{eqnarray} \label{sms}
I_{\rm v_\star <v_c} &=& \int_{0}^{v_c} dv_\star 4\pi f(v_\star) v_\star^2    \\
&=& \frac{2}{\sqrt{\pi}}\frac{\Gamma(\gamma+1)}{\Gamma(\gamma-1/2)} \int_{1/2}^1dx ~x^{\gamma-3/2}\sqrt{1-x} ~, \nonumber
\end{eqnarray} 
This function varies smoothly from $0$ when $\gamma=0.5$ to $0.5$ when $\gamma=2$.
For the fast moving stars an  (ad-hoc) approximation is
\begin{eqnarray}
I_{\rm v_\star >v_c}&=& \int^{\sqrt{-2 \phi(r)}}_{v_c} dv_\star 4\pi f(v_\star) v_\star^2    \left[ {\rm ln} \left( \frac{v_\star+v_c}{v_\star-v_c} \right)-2\frac{v_c}{v_\star} \right]
\nonumber
\\
&\simeq&0.1721+0.5280\gamma-0.1812\gamma^2  
-529.2\exp(-14.46\gamma)\label{fms}~. \nonumber
\end{eqnarray}
For   $\gamma >1$ , $I_{\rm v_\star >v_c}\sim0.53$, and when $\gamma\sim 2$ the contribution from the slow stars is of order   $\sim {\rm ln} \Lambda$ larger than 
the  frictional force due to the fast moving stars.

Given these expressions,  and setting $L=\sqrt{GM_{\bullet}r}$ in equation~(\ref{eqam})
we have
\begin{eqnarray}\label{am}
{dr\over dt}=-\frac{8\pi \sqrt{G} \tilde{\rho} \tilde{r} ^{\gamma}} 
{M_{\bullet}^{3/2}} 
 \times\left[{\rm ln \Lambda}~I_{\rm v_\star <v_{\rm c}} +I_{\rm v_\star >v_{\rm c}}\right]m_{\rm cl} r^{5/2-\gamma} ~.~~~~~~
\end{eqnarray}
This expression can be then used to  compute orbits for
both point-like and  extended objects~(e.g., a stellar cluster that experiences mass loss during inspiral) that move 
in a stellar cusp near a \sbh.
We identified our model with the center of a galaxy like M87. We
adopted  $M_{\bullet}= 3\times 10^9~{M_{\odot}}$~\citep{macchetto,GB11}, 
a core velocity dispersion  $\sigma_{h}=278 {\rm kms^{-1}}$~\citep{yea,L:92} and we used the relation 
$\sigma_h^2=4\pi G\rho_h(r_h/3)^2$ with $r_h=600~$pc to obtain the core density: $\rho_h=35~M_\odot{\rm pc^{-3}}$.
Taking $\gamma=0.5$, the density at  $\tilde{r}=r_{\rm infl}=300~$pc 
is $\tilde{\rho}=50M_\odot{\rm pc^{-3}}$. 

For a test particle of mass $m_{\rm cl}$, 
integrating equation~(\ref{am}) yields
\begin{eqnarray}\label{rt1}
r(t)&=&\Big[r_{in}^{\gamma-3/2}-\frac{8\pi \sqrt{G} \tilde{\rho}  \tilde{r} ^{\gamma}(\gamma-3/2) } 
{M_{\bullet}^{3/2}} \\
&& \times\left[{\rm ln \Lambda}~I_{\rm v_\star <v_{\rm c}} +I_{\rm v_\star >v_{\rm c}}\right]
m_{\rm cl}\times t\Big]^{\frac{1}{\gamma-3/2}}~,\nonumber
\end{eqnarray}
for $\gamma\ne3/2$, and
\begin{equation}\label{rt2}
r(t)=r_{in}\exp\left( -\frac{8\pi \sqrt{G}  \tilde{\rho} \tilde{r} ^{\gamma} }
{M_{\bullet}^{3/2}} [{\rm ln \Lambda}~I_{\rm v_\star <v_{\rm c}} +I_{\rm v_\star >v_{\rm c}}] m_{\rm cl}\times t \right)~
\end{equation}
for $\gamma=3/2$.

Equation~(\ref{rt1})  corresponds to a characteristic   decay time for angular momentum loss (for $\gamma>3/2$):
\begin{eqnarray}\label{taubh}
\tau_{\bullet}
&=&4\times 10^8~{\rm yr}\frac{\left[{\rm ln \Lambda}~I_{\rm v_\star <v_{\rm c}} +
I_{\rm v_\star >v_{\rm c}}\right]^{-1}}{\gamma-3/2}\\
&& \times M_{\bullet,9.5}^{3/2}\tilde{\rho}_{50}^{-1} \tilde{r}_ {300}^{-3/2}m_{{\rm cl},6}^{-1}\left(\frac{r_{in}}
{\tilde{r}}\right)^{\gamma-3/2} \nonumber
\end{eqnarray}
where $M_{\bullet,9.5}=M_{\bullet}/3\times10^9~M_{\odot}$, 
 $\tilde{\rho}_{50}=\tilde{\rho}/50~M_{\odot}~{\rm pc^{-3}}$,  and $\tilde{r}_ {300}=\tilde{r}/300~{\rm pc}$.
We note that, for $\gamma<3/2$, $\tau_{\bullet}$ becomes negative and, formally, equation~(\ref{am}) gives an infinite decay time to the center. In this latter case the dynamical friction timescale can be re-defined as the time required to reach a radius which is some fraction, $\zeta$, of its initial 
value: $\tau_{\bullet}\times\left(1-\zeta^{\gamma-3/2} \right)$, with $\tau_{\bullet}$ from equation~(\ref{taubh}).

In the case of extended objects, 
we assume that  the central properties of the globular cluster~(i.e. $\sigma_K$ and $r_K$)  remain unchanged during inspiral.
 The cluster's limiting radius  is given by equation~(\ref{rtr2}), which permits  expressing the satellite mass as a function of radius.
Setting $m_{\rm cl}=m_t$ in equation~(\ref{am}) and assuming that the galactic potential is dominated by the \sbh, we obtain
\begin{eqnarray}\label{rvst}
r(t)&=&\Big[r_{in}^{\gamma-3}-\frac{2^{3/2}\pi}{\sqrt{3}}\frac{\tilde{\rho}\tilde{r}^\gamma(\gamma-3)} 
{GM_{\bullet}^2} \\
&& \times[{\rm \ln \Lambda}~I_{\rm v_\star <v_{\rm c}} +I_{\rm v_\star >v_{\rm c}}]\sigma_K^3\times t\Big]^{\frac{1}{\gamma-3}}~,\nonumber
\end{eqnarray} 
and
\begin{eqnarray}\label{taubh2}
\tau_{\bullet}&=&5\times10^{9}~{\rm yr}
\frac{\left[{\rm ln \Lambda}~I_{\rm v_\star <v_{\rm c}} +I_{\rm v_\star >v_{\rm c}}\right]^{-1}}{3-\gamma}
\left(\zeta^{\gamma-3}-1\right)  \nonumber\\
&& \times M_{\bullet,9.5}^{2} \tilde{\rho}_{50}^{-1}\tilde{r}_{300}^{-3} \sigma_{K,10}^{-3} 
\left(\frac{r_{in}}
{\tilde{r}}\right)^{\gamma-3}~.
\end{eqnarray}

We used equations~(\ref{rt1}) and (\ref{rvst}) to trace the orbital evolution of massive stellar clusters 
in the core of our galaxy model.
We take a limiting  value for the mass of $m_{K}=4\times10^6 ~M_{\odot}$, corresponding to  the non-truncated 
model, when the cluster is far from the center.
The simulated orbits~(upper panel of Figure~\ref{Fig4}) demonstrate that the dynamical friction time-scale  in the core of  bright elliptical galaxies
is extremely  long. Also very massive clusters~($\sigma_K\gtrsim 30~{\rm km~s^{-1}}$)
starting well within the galaxy core~($r_{in}\sim 300~$pc), 
do not reach the center  after one Hubble time.

In the lower panel of Figure~\ref{Fig4}, we use equations~(\ref{rtr2}) and (\ref{deltarho})
to compute the density profile of a \NSC\ resulting from the accumulation of
 $500$ equally massive globular clusters 
in the center of our M87 galaxy model. 
At radii smaller than $r_{\rm disr}$ (marked by black arrows in the plot) the clusters are fully disrupted by the interaction
with the \sbh\ and  our
integrations are no longer valid.  $N$-body simulations show that the density profile of stars
will be very flat or even declining  toward the \sbh\  inside $r_{\rm disr}$, as the stars accumulate near
the radius of disruption~\citep{Sungsoo03,FU09,FU10}.

In order to hightligh 
the role of the central \sbh\ in determining the structure of the growing nucleus 
 Figure~(\ref{Fig4}) also shows the density profile of the nucleus when the mass of the \sbh\ is set
to zero. In this case, most of the cluster mass is delivered in the inner $\sim 100~$pc and consequently the 
 \NSC\ is much more centrally concentrated. After $500$ inspiral events, 
the \NSC\ appears to be distinct from the galaxy background density profile.
Apparently, \NSC s assembled around \sbh s have a spatially more diffuse configuration and lower densities than
those forming in galaxies with no \sbh .
We conclude   that  the formation of \NSC s in giant ellipticals 
might be  inhibited by the presence of their central \sbh \ for two basic reasons:
(i)~the long dynamical friction timescale of massive objects in the galaxy core;
(ii)~clusters are disrupted by the strong tidal field of the \sbh\ producing a merger remnant with density profile that
rises only modestly above that of the background galaxy.

We note in passing that
the latter argument  may also apply  to the case in which a
giant elliptical  galaxy accretes a smaller
galaxy containing a dense central nucleus. 
The large cores observed in the central light profile 
of bright ellipticals are usually interpreted in terms of 
 the  scouring effect  of  \sbh\ binaries forming during major merger events~\citep{MM}. 
 It is less clear however how such cores can be preserved up to the present epoch
despite the large number of minor mergers  that are predicted to occur 
by standard cosmological models. For instance, an M87 like galaxy is expected
to have accreted few galaxies of the size of the MW in the last $\sim 5~$Gyr~\citep{FMB}.
A plausible explanation  for the lack of central dense stellar concentrations
in the brightest galaxies was provided by~\citet{MerrittCruz2001}.
These authors performed  high resolution $N$-body simulations 
 of the accretion of high-density dwarf galaxies by low-density giant galaxies.
They found that the cusp of the secondary galaxy  is disrupted during the merger
by the giant galaxy \sbh\ tidal field, producing a remnant
with a central density that is only slightly higher than that of the giant galaxy initially;
removing the black hole from the giant galaxy allowed the smaller galaxy to 
remain essentially intact and led to the formation of a high central density cusp, 
contrary to what is observed.

\section{Discussion}
\subsection{Long term evolution of \NSC s around \sbh s}\label{mbh}
In the previous sections we showed that \NSC s resulting
from the accumulation of globular clusters around  \sbh s  have lower central densities and larger
cores than those  forming in galaxies without \sbh s.  Such a result is consistent with the results 
obtained  using more sophisticated $N-$body simulations~\citep{AM12-2}. 

Our analysis however did not account for the effects of internal dynamics occurring in the \NSC\ during and after its formation. 
In a preexisting \NSC , the presence of a \sbh\ would inhibit core collapse,
causing instead the formation of a \citet{BW:76} cusp on the two-body relaxation 
timescale, followed by a slow expansion as stars are tidally disrupted~\citep{M09}.
Whether or not a Bahcall-Wolf cusp  will form depends on the initial core size of the nucleus 
and on its relaxation time.

For nuclei  similar to the MW \NSC , relaxation times
are  too long to assume that they have reached such a collisionally relaxed state, but 
 they are still sufficiently short  that gravitational encounters would substantially affect their 
structure over the age of the galaxy. Two body gravitational interactions will 
cause the central density of the nucleus to increase and the core to shrink 
as the stellar  distribution evolves  toward  the Bahcall-Wolf form~\citep[e.g.,][]{preto}.
The time evolution of the core radius can be approximately described by the expression~\citep{AM12-2}
\begin{equation}\label{evrel}
r_{\rm core}(t)=1.57~{\rm pc} \exp\left[ t/0.25t_{\rm relax} \right]	~,
\end{equation}
where $t_{\rm relax}$ is the relaxation time computed at the radius of influence of the \sbh .
For the MW, $t_{\rm relax}\sim 20 ~$Gyr and $M_{\bullet}\approx 4\times10^6~M_{\odot}$.
A black hole of this mass   corresponds to
 a core of  $\sim 3~$pc~(Figure~1). From equation~(\ref{evrel}), we see that 
 gravitational encounters occurring during and after the formation of the \NSC\ 
 will reduce the core size to   $\sim 1~$pc after $10^{10}~$yr. 
Such a value is more consistent with the size of the  core  observed 
in the distribution of late-type (old) stars in the inner parsec of the MW~\citep{Buch,Do,Bartko,YZ,Nadeen}.
Whether \NSC s in other ``power-law'' galaxies  will turn out to have stellar cores similar 
to that of the MW remains to be seen.

\subsection{The distribution of stellar remnants near \sbh s}
\NSC s containing \sbh s are  expected to be the main birthplace of 
  gravitational wave~(GW) sources for  space-based interferometers~\citep{HU:03}. These include
 the capture of stellar-mass black holes~(BHs) by \sbh s, also  called 
 ``extreme mass-ratio inspirals''~\citep[EMRIs;][]{M11,AS12}.
Understanding systems like the MW NSC and their origin 
is therefore crucial for making predictions about
the event rates for low frequency gravitational wave detectors. 

Most of the  EMRI rate calculations reported in the  literature are derived  under the assumption
that  the galactic nucleus had enough time to reach a state of mass segregation, which implies  a high density of 
BHs near the center~\citep[e.g.][]{F06,HA06,AH:09}.  This appears to be in conflict  with observations
 which suggest an unrelaxed state for the distribution of stars at the GC. 
\citet{M10} and \citet{AM12} demonstrated  that in the absence of an initial  cusp in the stars,
even  after $5-10$~Gyr the density of BHs  could remain substantially below the densities inferred from steady-state models.  

All these previous studies, however, assumed 
that the stellar BHs had the same phase-space distribution initially
as the stars. This is most likely to be a poor assumption
if a  substantial fraction of the \NSC \ mass comes from orbitally decayed globular clusters.
In the merger model the resulting distribution of stellar remnants will
reflect their distribution in their parent  clusters just before they 
reach the center of the galaxy. 

In a dense stellar cluster, BHs formed by the supernova explosions of the most massive stars 
tend to segregate into the cluster core and form a sub-cluster of BHs which  dynamically 
decouples from the rest of the cluster~\citep{Spitzer}.
When the central density of BHs becomes large enough, BH-BH binary formation
becomes efficient. Subsequent dynamical interactions involving binary BHs and higher multiplicity 
systems will tend to eject the BHs from the cluster until only a few of them are 
left~\citep[e.g.,][]{DBGS10,DBGS11,BBK10}. 

We may consider two possibilities \footnote{We make here the reasonable assumption
that the timescale for a clusters  to reach the center of the galaxy is long 
compared to the mass-segregation timescale of their BH population.}: (i) the 
 dynamical friction timescale of star clusters is short compared to the 
encounter driven evaporation timescale of their BH sub-cluster.  
In this case, the mass-segregated BH sub-clusters, due to their high central densities,  can reach 
galactocentric radii as small as $\sim 0.1~$pc~\citep{ANT12}. The preferential removal of stars
 from the outer parts of the clusters by the strong galactic tidal field 
  might lead to the formation of a \NSC \ with 
    a central over-abundance of BHs when compared  to predictions from  standard mass functions~\citep{BK11}.
 (ii) The  orbital  decay  timescale   is 
  shorter than the evaporation timescale of the cluster BH sub-system.  
Most of the BHs will be dynamically ejected in this case before the
 cluster  reaches the center of the galaxy. As a consequence of this, very few BHs
 are delivered to the vicinity of the \sbh .
 
All of the above mentioned topics require a dedicated study which 
we defer to a future paper.

\subsection{Dissipative \NSC~ formation}
Our work shows that a purely dissipationless  merger scenario can explain, without obvious difficulties, 
the basic physical properties of \NSC s.
It is important to note that  we adopted a rather idealized model of an isolated galactic
spheroid~(with/without \sbh). This idealized model enabled us to neglect
some  gradients of the role of galaxy  merging in  \NSC\ formation
which  can only be  addressed by means of comprehensive cosmological
hydrodynamical simulations.
Simulations demonstrate that tidal torques in major mergers of gas-rich galaxies 
can induce rapid inflow of   gas  into the center of a galaxy
followed by intense nuclear starbursts  which results
 in a central light ``excess''  in the surface brightness profile of the
galactic spheroid~\citep[e.g.,][]{Hopkins+2008,Hopkins+2009}.
Unfortunately,  the spatial resolution~($\gtrsim 50~$pc) of current  cosmological simulations
 is not high enough to address the role of galaxy mergers in the context of \NSC s, at least for
low and  low-intermediate mass galaxies. 
In addition, it is not clear whether this picture would  apply to elliptical, early-type  galaxies which 
lack the large gas reservoirs of spirals, and thus should not experience frequent central starbursts.

On the other hand, the fact that the nuclei formation 
 histories were in part  governed by local and dissipative factors 
is supported by a wide range of observational phenomena.
As an example, the gas distribution and the kinematics of the gas in the nearby spiral NGC6946 suggest
the presence  of a central, small-scale, S-shaped stellar bar
which  appears to  funnel gas towards the galaxy nucleus~(within the inner $\sim 10~$pc) where about 
$10^7~M_{\odot}$ of molecular gas have been accumulated.
Star forming events triggered by 
 the rapid inflow of gas in the center of the galaxy may then contribute  to the growth 
of its \NSC~\citep{S06}.
Support to a dissipative origin is added by the  fact that \NSC s
in general tend to have a wide range of stellar ages, including young stellar populations~\citep{Rossa,SETH06}.
Our GC contains for instance a large population of young massive stars that most likely formed locally 
following the infall and fragmentation of a dense gaseous clump in the vicinity of Sgr~A*~\citep{paumard+2006}. Such bursts may occur
continuously over the age of the Galaxy and thus produce a significant fraction  of its \NSC\  mass. 
Stellar population synthesis studies also show
 that the GC  appears to have undergone continuous and recurrent star formation
over the last $10~$Gyr, but
it is not possible to fit the observations with ancient burst models, such as would be appropriate for an old population
of stars that originated  in globular clusters~\citep{Figer2004}.   

Previous work incorrectly used the latter arguments to argue against a dissipationless origin
for  \NSC s~\citep[e.g.,][]{Milos04,Nay09}.
It is important to stress that although observations probe recent  and episodic star formation they
do not exclude that the bulk 
of the GC stellar population  is in old stars.
In fact, the GC luminosity function appears to be consistent with 
a star formation history in which  a large fraction (about 1/2) of the mass consists 
of old ($\sim 10$~Gyr) stars and the remainder is due to  continuous star formation~\citep{AM12-2}. 
Accordingly, \citet{Ol11}  found that about 80 per cent of the stellar mass in the inner parsec of the Galaxy is in old stars 
that formed more than $5~$Gyr ago. 
We add that, although globular clusters  in our Galaxy are exclusively old stellar 
systems~\citep[e.g.,][]{Ros},
this might not be always the case in other  galaxies.
It is therefore not clear what fraction of the mass in \NSC s is 
contributed by local gaseous fragmentation.
More investigation needs to be done in order to explore  the 
implications of in-situ star formation as the origin of galactic nuclei.

\subsection{\NSC s morphology and kinematics}
The morphology and kinematics of \NSC s are of great importance for understanding 
their origin~\citep[e.g.,][]{delorenzi+12}.

Aspherical \NSC s are commonly observed in external galaxies. \citet{SETH06} found  that the three edge-on late-type galaxies 
IC~5052, NGC~4206 and NGC~4244 have  nuclei that are strongly elongated along the 
plane of their host galaxies disks. Such clusters show evidence for multiple morphological components, 
with a disk-like young stellar population superimposed on an older more spherical component.
The  radial velocity map of the  nucleus  in NGC~4244, the nearest of these three galaxies~($D=4.1~$Mpc),
shows evidence for  strong rotation,
$30~{\rm km~s^{-1}}$ at $10~$pc from the center, compared to a central velocity dispersion
of $\sim 28~{\rm km~s^{-1}}$~\citep{seth}.   
There is also evidence for flattening and rotation in the M33 nucleus~\citep{L:98,MA:99}. The M33 \NSC \ is 
elongated along the major-axis of the galaxy  and rotates at $\sim 8~{\rm km~s^{-1}}$, while the central velocity dispersion 
is $\sim 27~{\rm km~s^{-1}}$~\citep{GE:01}.
The MW \NSC \ also appears to be rotating parallel  to the overall Galactic rotation~\citep{trippe,S09}.
Unfortunately, as a consequence of the  strong interstellar extinction  along the GC line of sight,
our knowledge of the Galactic NSC morphology and size remains  very limited.

Is the  observed  rotation of \NSC s
consistent with the predictions of a dissipationless model for \NSC\ formation?
Of course if a large fraction of the \NSC\ mass was formed via accretion of 
clusters  isotropically distributed   throughout the galaxy, no net rotation would be expected, since there will not be any
preferred  direction for inspiral.  If   the primary formation process is instead
 gas accretion from the galactic disk, this would  naturally 
explain both the flattening and the fact that \NSC s rotate parallel to their host  galaxy rotation.

Although young and rotating components of \NSC s 
might be difficult to reconcile with a globular cluster origin (due to the long time scale 
for inspiral compared to the ages of the stars),
we believe that  the evidence for \NSC \ rotation alone, at least in late-type galaxies, is not 
 a strong argument against a dissipationless  origin.
For instance, clusters falling into the GC could have originated in the 
inner part of the Galactic disk  and they will therefore share in its rotation~\citep{Hartmann}. Another
 possibility  is that globular clusters  crossing  the Galactic disk  experience a greater frictional force from
 the increased local stellar/gas density~\citep{B:10} and hence they can be dragged down into the disk plane and  transported into the central region of the galaxy where they then accumulate to form a dense nucleus. 
 In either case, the forming \NSC\ will appear to rotate in the same sense of the Galaxy.
These  hypotheses  lead to basic predictions  that might be testable with future observations; specifically:
(i)~\NSC s   in early-type galaxies,  which do not have extended massive disk-like structures,
might have slower rotation with respect to \NSC s  of spiral galaxies.
(ii)~There should be some mild depletion of stellar clusters in the Galactic disk as compared to off the disk 
plane~\footnote{I am thankful  to A.~Madigan for pointing this out.}.
  
In conclusion, comparing  kinematic data with 
simulations for distinguishing between gas and cluster accretion may be
difficult, since the detailed structure  of simulated \NSC s  varies in an important way
with the orbital  initial configuration of the infalling clusters. 
Future observational work  should be able to provide
more reliable models for the spatial distribution of globular clusters near 
the center of galaxies, where timescales for infall are reasonably short.

\subsection{The lack of \NSC s in faint galaxies}
Galaxies fainter than $M_B\sim -12$ do not contain prominent stellar nuclei~\citep{VdB}.
The lack of \NSC s in dwarf spheroidals  was explained in the context of the merger formation scenario 
in \citet{Goerdt}. These authors used high resolution $N$-body simulations
to study the orbital decay of globular clusters in  dark matter halos 
with a core-density structure in the central regions, similar to what inferred by
observations~\citep{PK:90}.
In these models the standard Chandrasekhar's dynamical friction formula breaks down~\citep{TW1984,w:86} and
one finds that a massive object stalls at roughly the dark matter core radius~\citep[e.g.,][]{Read+06,Inoue09}; the number of globular clusters predicted to inspiral would be then $\lesssim 1$.
This suggests that the lack of \NSC s in the fainter spheroidals  could be simply a
consequence  of the arbitrarily  long sinking timescales of stellar clusters in these systems.

Alternatively, the lack of nuclei in
faint galaxies could be related to 
the small total number, or even absence, of globular clusters in these galaxies.
In the Local Group,  globular clusters  seem in fact
 to disappear in stellar spheroids fainter than $M_B \sim -12.5$~\citep[e.g.,][]{Peng2008},
which would make the formation of \NSC s through globular cluster merging impossible
in these systems~\citep[see also][]{turner+12}.

\section{Conclusions}
In this paper we considered a dissipationless formation model for \NSC s
where globular clusters orbital decay and merge at the center  of a galaxy to form a compact nucleus.
Our main results are summarized below.

\begin{itemize}
\item[1]  The observed scaling relation between \NSC\ masses  and the velocity dispersion
of their host spheroids, the $M_{\rm NSC}-\sigma$ relation, 
is difficult to reconcile with a purely dissipative formation
model for the nuclei. These models predict that  $M_{\rm NSC}$
is a steeply rising function of $\sigma$. The observed $M_{\rm NSC}-\sigma$ relation is instead
in agreement with  the predictions of a dissipationless formation model.
Dissipationless formation modes produce relations that are substantially 
shallower than the corresponding \sbh \ 
scaling relations and are therefore more consistent with observations.

\item[2]The globular cluster merger model, in the absence of a central \sbh ,
naturally reproduces the observed relation between 
the size of galactic nuclei and their total luminosity, 
$R\propto \sqrt{\mathcal{L}_{\rm NSC}}$. When a \sbh \ is present,
the dependence of the \NSC \ radius on its mass is substantially weaker 
than the observed relation because  the size of the \NSC \  is mainly determined by the fixed tidal field of the \sbh .

\item[3] We derived explicit  expressions for the
orbits of globular clusters owing to
 dynamical friction and subject to mass-loss 
  due to their tidal interaction with the galaxy~(equation~\ref{sfc2})
  or with a central \sbh ~(equation~\ref{rvst}). 
These expressions were used  to (i)~address the possibility that
\NSC s in galaxies similar to the MW could have been assembled  
via cluster migration and mergers; and (ii)~to follow the orbital evolution 
of globular clusters in the core of  bright elliptical galaxies.

\item[4] 
 \NSC s that form  through the mergers of globular clusters 
have a density profile characterized by a parsec-scale core and an envelope that falls off as $\rho\sim r^{-2}$.
These properties are similar to those of the MW \NSC . 
A \NSC \ with mass comparable to that of the MW \NSC\ is obtained by 
assuming an initial total mass in stellar clusters which is consistent  with
the cluster formation efficiency inferred from observational studies of embedded
clusters in Galactic molecular clouds~(Figure~\ref{Fig1}).

\item[5] 
A pre-existing \sbh\ at the center of the stellar spheroid has a strong impact on the structure of the 
growing stellar nucleus~(Figure~\ref{Fig2}).  
Tidal stresses from a \sbh\  disrupt the clusters when
they pass within  their tidal disruption radius, $r_{\rm disr}$, 
limiting the density within that radius. 
Hence,  the density profile of the resulting \NSC\ will also have  core  size of $\sim r_{\rm disr}$. 
Removing the \sbh\ from the galaxy
 allows the stellar clusters to keep their inner structure almost unchanged 
 leading to the formation of a \NSC\ with higher peak  densities. We find that
 separating  the contribution of the nucleus 
from that of the galaxy could more difficult in
 stellar spheroids  with  more massive \sbh s.

 \item[6] 
For  globular clusters orbiting in the core of a massive elliptical galaxy like M87, the timescale 
 to reach the center   is much longer than one Hubble time~(Figure~\ref{Fig4}). 
 The essential  reason for this is that the phase-space density of a  shallow density cusp
of stars  around a \sbh \ falls to zero  at low energies: inside the galaxy core there are no
stars at any radius that move slower than the local circular velocity.
 The standard Chandrasekhar's formula
predicts little or even no frictional force in this case.
In addition, when a \sbh\ is included, the tidal field near $r_{\rm infl}$ becomes much stronger, 
resulting in faster mass loss and fewer stars deposited to the center.  
Based on these facts,
 we conclude the presence of central \sbh s  might inhibit  the  formation of compact nuclei in the brightest galaxies, 
 in agreement with what is observed.

\end{itemize}

 \bigskip
I am grateful to D.~Merritt for numerous discussions about the ideas presented in this paper
and for his detailed comments to an earlier version of this manuscript.
I  thank R.~Capuzzo-Dolcetta for encouraging me to study this problem,
 D.~Kruijssen and A.~Madigan for useful discussions and M.~Alvarez and
 A.~Graham  for their comments to an earlier version of this paper. 
 This material is based upon work supported in part by the National Science Foundation Grant No. 1066293 
and the hospitality of the Aspen Center for Physics.

\end{document}